  \documentstyle[12pt]{article}
    \def\l{\lambda}
    \def\m{\mu}

\begin{document}
  \begin{center}
    {\Large \bf Investigations of the broken SU(3) symmetry in
    deformed even-even nuclei}
  \end{center}
  \bigskip
  \begin{center}
    N.  Minkov, S.  Drenska, P.  Raychev, R.  Roussev \\
    {\em Institute for Nuclear Research and Nuclear Energy,} \\
    {\em 72 Tzarigrad Road, 1784 Sofia, Bulgaria}\\
\medskip
    Dennis Bonatsos \\
    {\em Institute of Nuclear Physics, N.C.S.R. ``Demokritos''}\\
    {\em GR-15310 Aghia Paraskevi, Attiki, Greece}
\end{center}

    \bigskip

    \begin{abstract}
    A collective vector-boson model with broken SU(3) symmetry is applied to
    several deformed even--even nuclei.  The model description of ground and
    $\gamma$ bands together with the corresponding B(E2) transition
    probabilities is investigated within a broad range of SU(3) irreducible
    representations (irreps) $(\l ,\m )$.  The SU(3)--symmetry characteristics
    of rotational nuclei are analyzed in terms of the bandmixing interactions.
    \end{abstract}

    The vector-boson model is founded on the supposition that
    the SU(3) symmetry is inherent for the well deformed even--even nuclei, so
    that the low--lying collective states of these nuclei could
    be united into one or several SU(3) multiplets, labeled by the
    irreps ($\l ,\m$) \cite{p:descr,a:over,p:matr}.
    The collective rotational Hamiltonian reduces this symmetry to the
    rotational group O(3) and thus the energy spectrum of the nucleus is
    generated. The basis states corresponding to the  reduction $SU(3)\supset 
O(3)$ are constructed with the use of two distinct kinds of vector
    bosons, whose creation operators $\mbox{\boldmath $\xi^{+}$}$ and
    $\mbox{\boldmath $\eta^{+}$}$ are O(3) vectors and in addition transform
    according to two independent SU(3) irreps of the type $({\l},{\m})=(1,0)$.
    The SU(3)--symmetry breaking Hamiltonian has the following form:
    \begin{equation}
    \label{eq:v}
    V=g_{1}L^{2}+g_{2}L\cdot Q\cdot L +g_{3}A^{+}A\ ,
    \end{equation}
    where $g_{1}$, $g_{2}$ and $g_{3}$ are the parameters of the model;
    $L$ and $Q$ are the angular momentum and quadrupole operators
    respectively;
    the term $A^{+}A$ is constructed by using the operator
    $A^{+}=\mbox{\boldmath $\xi^{+}$}^{2}\mbox{\boldmath $\eta^{+}$}^{2}-
    (\mbox{\boldmath $\xi^{+}$}\cdot\mbox{\boldmath $\eta^{+}$})^{2}$.
    In this model the ground state band
 (gsb) and the lowest $\gamma$ band belong to the same SU(3) irrep ($\l ,\m$).

    In the present work we study the global SU(3) characteristics of
    rotational nuclei.  We suppose that for a given nucleus the physically
    significant features of SU(3)--symmetry should be sought in certain
    regions of ($\l ,\m$) irreps instead of a single fixed irrep.  It is
    therefore of interest to study whether the available experimental
    information on the energy levels and transition probabilities could be
    used to estimate the SU(3) symmetry properties of the nucleus, in
    particular to outline the physically favored regions in the ($\l ,\m $)
    plane.  Then the rotational nuclei could be systematized accordingly.  In
    addition one can investigate the principal limits of 
applicability of the    SU(3) symmetry in nuclei.

    In order to implement these investigations, we have elaborated the model
 scheme for calculations in arbitrary high irreps with $\m \geq2$ \cite{MSU3}.
    We have considered eight rare earth nuclei ($^{164}$Dy, $^{164-168}$Er,
    $^{168,172}$Yb, $^{176,178}$Hf) and one actinide nucleus ($^{238}$U) for
    which the model descriptions of the gsb and $\gamma$--band energy levels
    and the concomitant B(E2) transition ratios have been evaluated [in the
    form of root mean square (RMS) fits] in SU(3) irreps within the range
    $10\leq\l\leq 160$ and $2\leq\m\leq 8$. These nuclei represent regions of
    SU(3) spectra with different magnitudes of energy splitting between the
gsb and the first $\gamma$-band. As a measure of the splitting we use the ratio
    $\Delta E_{2}=(E_{2_{2}^{+}}-E_{2_{1}^{+}})/E_{2_{1}^{+}}$,
    where $E_{2_{1}^{+}}$ and $E_{2_{2}^{+}}$ are the experimentally measured
$2^{+}$ energy levels, belonging to the gsb and the $\gamma$-band respectively.
In the rare earth region this ratio varies within the limits $7\leq\Delta 
E_{2}\leq 18$, while in the actinides one
    observes values in the range $13\leq\Delta E_{2}\leq 25$.

    For the nuclei $^{164-168}$Er, $^{164}$Dy and $^{168}$Yb one observes
    small band splitting ratios $\Delta E_{2}\sim 8-10$.  In the typical case
    of $^{168}$Er ($\Delta E_{2}=9.3$) the model calculations are implemented
    in the SU(3)--irreps within the range $10\leq\l\leq 90$ and $\m =2,4,6,8$.
    The results obtained for the description of the energy levels are shown in
    Fig.  1(a), where the standard RMS-factor $\sigma_{E}$ is given as a
    function of the quantum number $\l$.  One finds that for $^{168}$Er the
    model scheme provides a clearly outlined region of ``favored'' multiplets
    in the $(\l ,\m )$--plane, including relatively small $\l$--values
    $\l =14-20$ and $\m =2,4,6$.  Outside
    this region $\sigma_{E}$ increases gradually with the increase of $\l$ and
    for $\l >40$ it saturates towards the values obtained in the $(\l
    ,8)$--multiplets.  It is also clear that the best description of the
    energy levels corresponds to the multiplet $(20,2)$, which provides the
    absolute $\sigma_{E}$--minimum observed in the considered variety of $(\l
    ,\m)$--multiplets.  Almost the same picture has been
    obtained in the other nuclei with small SU(3) energy splittings (see
    Table~1).    Since the favored $(\l ,\m )$ regions are
    determined on the basis of the experimental gsb and $\gamma$--band
    characteristics, the above result can be interpreted as a natural physical
    signature of the broken SU(3) symmetry in these nuclei.

    \begin{table}
 \caption{The parameters of the fits of the energy levels and the  transition ratios \protect\cite{MSU3}
    of the nuclei investigated are listed for the
    $(\l ,\m )$ multiplets which provide the best model descriptions.  The
    Hamiltonian parameters $g_{1}$, $g_{2}$ and $g_{3}$ are given in keV.  The
    quantities $\sigma_{E}$ (in keV) and $\sigma_{B}$ (dimensionless)
    represent the energy and the transition RMS factors respectively
    The splitting ratios $\Delta E_{2}$
    (dimensionless) are also given.}

    \bigskip
    {\small
    \begin{center}
    \begin{tabular}{cccccccc}
    \rule{0em}{2.2ex}
 & & & & & & &  \\
\hline\hline
Nucl&$\Delta E_{2}
$&$\l ,\m$&$\sigma_{E}$&$\sigma_{B}$&$g_{1}$&$g_{2}$&$g_{3}$ \\
\hline\hline
${}^{164}\rm Dy
$&$9.4$&$16, 2$&$14.1$&$0.52$&$-1.159$&$-0.321$&$-0.590$  \\
${}^{164}\rm Er
$&$8.4$&$18, 2$&$8.1$&$0.14$&$3.625$&$-0.238$&$-0.513$  \\
${}^{166}\rm Er
$&$8.8$&$16, 2$&$5.8$&$0.47$&$2.942$&$-0.235$&$-0.572$  \\
${}^{168}\rm Er
$&$9.3$&$20, 2$&$3.2$&$0.28$&$4.000$&$-0.181$&$-0.401$  \\
${}^{168}\rm Yb
$&$10.2$&$20, 2$&$7.9$&$0.27$&$0.500$&$-0.271$&$-0.501$  \\
${}^{172}\rm Yb
$&$17.6$&$\geq 80, 2$&$6.8$&$0.12$&$9.875$&$-0.017$&$-0.052$  \\
${}^{176}\rm Hf
$&$14.2$&$\geq 70, 2$&$15.0$&$0.17$&$9.547$&$-0.030$&$-0.062$  \\
${}^{178}\rm Hf
$&$11.6$&$34, 2$&$7.0$&$0.86$&$8.322$&$-0.083$&$-0.213$  \\
${}^{238}\rm U
$&$22.6$&$\geq 60, 2$&$1.6$&$0.08$&$-37.697$&$-0.360$&$-0.098$ \\
    \hline\hline
    \end{tabular}
    \end{center}
  }
    \label{tab:favor}
    \end{table}

    The nuclei $^{172}$Yb, $^{176}$Hf and $^{238}$U are characterized by large
    $\Delta E_{2}>14-15$ values.  As a typical example consider the $^{172}$Yb
    case where $\Delta E_{2}=17.6$.  In Fig.  1(b) the RMS factors
    $\sigma_{E}$ obtained for this nucleus are given for the $(\l ,\m
    )$--multiplets in the range $10\leq\l\leq 160$ and $\m =2,4,6$.  Here we
    find an essentially different picture.  In the small $\l$'s the
    $\sigma_{E}$--factor decreases with increasing $\l$ and further at large
    $\l\sim 60-80$ saturates gradually to a constant value $\sigma_{E}\sim
    6.5$ keV without reaching any minimum.  A similar picture is observed in
    $^{176}$Hf and $^{238}U$ \cite{MSU3}.  Thus for the nuclei with large
    bandsplitting the calculations indicate the presence of a wide lower limit
    of the quantum number $\l$ ($\l >60-80$), instead of a narrow region of
    favored multiplets.  For the nucleus $^{178}$Hf (with a medium energy
    splitting $\Delta E_{2}= 11.6$) one finds a slightly expressed
    $\sigma_{E}$--minimum, disposed in the medium $\l$--region $30\leq\l\leq
    40$ \cite{MSU3}.  This result suggests the presence of a smooth transition
    from the picture in Fig1(a) to the one in Fig1(b).

    \begin{figure}
    \vspace{3in}
    \caption{
    The energy RMS factor $\sigma_{E}$, obtained for the nuclei $^{168}$Er
    (part (a)) and $^{172}$Yb (part (b)), are plotted as a function of the
    quantum number $\l$ at $\m =2$ (circlets), $\m =4$ (squares), $\m =6$
    (triangles), and $\m =8$ (asterisks).}
    \end{figure}

    The obtained results can be analyzed in terms of the band-mixing
    interactions \cite{MSU3}.  The estimates of the Hamiltonian matrix
    elements show that the increase in the quantum number $\l$ is connected
    with the corresponding decrease in the mixing interaction between the gsb
    and the $\gamma$--band within the framework of the SU(3) symmetry.  Hence
    for the nuclei with small band splitting ($^{164}$Dy, $^{164-168}$Er,
    $^{168}$Yb) the relatively small $\l$--values ($\l\sim 16-20$) indicate
    that the gsb and the $\gamma$--bands are strongly mixed.  In the nuclei
    with a large band splitting ($^{172}$Yb, $^{176}$Hf, $^{238}$U) the large
    $\l$'s correspond to a weak interaction between the two bands. This means
    that for the latter nuclei the rotational character of the gsb and the
    $\gamma$ bands should be better developed.  Indeed the case of the nucleus
    $^{238}$U with a very large splitting ratio $\Delta E_{2} =22.6$ and a
    well pronounced rotational structure of the gsb supports the above
    supposition.  Further analysis shows that the large $\l$ limit
    ($\l\rightarrow\infty $) corresponds to an asymptotical decrease of the
    band interaction to zero.  Thereby the multiplet splits into distinct
    noninteracting rotational bands and the SU(3) symmetry gradually
    disappears.  This situation is equivalent to the group contraction process
    in which the SU(3) algebra reduces to the algebra of  the triaxial rotor 
group T$_{5}\wedge$SO(3).

    The implemented investigations allow us to conclude that the violation of
    the SU(3) symmetry, measured by the splitting ratio $\Delta E_{2}$
    determines to a great extent the most important SU(3) properties of
    deformed nuclei.  On this basis we suggest that the strongly split spectra
    should be considered as special cases in which the gsb and the
    $\gamma$--bands are weakly coupled.  This allows the possibility for
    transition from the gsb--$\gamma$ band coupling scheme (in the nuclei with
    small $\Delta E_{2}$) to an alternative collective scheme (in the cases of
    large $\Delta E_{2}$), in which the gsb is situated in a separate irrep.
    In other words the broken SU(3) scheme is favored in the case of weak
    $2^+$ splitting, while strong $2^+$ splitting favors SU(3) schemes like
    the one of the Interacting Boson Model \cite{Arima3} (IBM), in which the
    gsb is situated in a separate irrep.  The  collective dynamical mechanism 
causing such a transition from the broken
    SU(3) of the present model to the pure SU(3) of the IBM could be sought in
    the framework of the more general dynamical symmetry group Sp(6,$\Re$).

    This work has been supported in part by BNFSR under contracts no F--547
    and no F--415.  One of the authors (DB) has been supported by the EU under
    contract ERBCHBGCT930467 and by the Greek General Secretariat of Research
    and Technology under contract PENED95/1981.

    \end{document}